\documentclass{elsart}
\usepackage{epsfig}
\usepackage[dvips]{color}
\newcommand{\btbl}{\begin{tabular}}
\newcommand{\etbl}{\end{tabular}}
\newcommand{\beq}{\begin{displaymath}}
\newcommand{\eeq}{\end{displaymath}}
\newcommand{\ben}{\begin{equation}}
\newcommand{\een}{\end{equation}}
\newcommand{\bea}{\begin{eqnarray}}
\newcommand{\eea}{\end{eqnarray}}
\def\bar{\begin{array}}
\def\ear{\end{array}}

\begin{document}
\begin{frontmatter}

\title{Escape Times in Fluctuating Metastable Potential and Acceleration
of Diffusion in Periodic Fluctuating Potentials}

\author{Bernardo Spagnolo\corauthref{cor}},
 \corauth[cor]{Corresponding author: http://gip.dft.unipa.it;}
 \ead{spagnolo@unipa.it}
 \author{ Alexander A. Dubkov$^{\circ}$, Nikolay V. Agudov$^{\circ}$}
\address{$^{\ast}$INFM-Group of Interdisciplinary Physics and  Dipartimento di
Fisica e Tecnologie Relative, Universit\`a di Palermo, Viale delle
Scienze - 90128 Palermo, Italy}
\address{$^\circ$Radiophysics Department, Nizhni Novgorod State
University, 23 Gagarin Ave., Nizhni Novgorod, 603950 Russia}

\begin{abstract}
The problems of escape from metastable state in randomly flipping
potential and of diffusion in fast fluctuating periodic potentials
are considered. For the overdamped Brownian particle moving in a
piecewise linear dichotomously fluctuating metastable potential we
obtain the mean first-passage time (MFPT) as a function of the
potential parameters, the noise intensity and the mean rate of
switchings of the dichotomous noise. We find noise enhanced
stability (NES) phenomenon in the system investigated and the
parameter region of the fluctuating potential where the effect can
be observed. For the diffusion  of the overdamped Brownian
particle in a fast fluctuating symmetric periodic potential we
obtain that the effective diffusion coefficient depends on the
mean first-passage time, as discovered for fixed periodic
potential. The effective diffusion coefficients for sawtooth,
sinusoidal and piecewise parabolic potentials are calculated in
closed analytical form.

\end{abstract}
\end{frontmatter}

\section{Mean first-passage times in dichotomously fluctuating metastable potential}

The problem of escape from metastable state in fluctuating
potential is of great importance to many natural systems, ranging
from physical and chemical systems to biological complex systems.
The characteristic feature of all these complex nonequilibrium
systems is that they are open systems, which are in contact with a
fluctuating environment. A typical problem is the enhancement of
stability of metastable states in fluctuating potentials due to
the external noise \cite{Man,Agu,Fia,Day,Dub}. The noise enhanced
stability (NES) phenomenon was observed experimentally and
numerically in various physical systems. This effect implies that
a system remains in the metastable state for a longer time then in
deterministic case and mean first-passage time (MFPT) has a
maximum at some noise intensity. Specifically NES phenomenon was
found in the transient dynamics of a periodically driven
overdamped particle in a noisy cubic potential \cite{Day}, in a
tunnel diode by investigating the escape time in a periodically
driven metastable system in a strong forcing regime \cite{Man}.
More recently it was found that several properties of the
noise-induced stability in one-dimensional maps
 and in fluctuating bistable potentials are
related to NES effect \cite{Wac}. The noise induced slowing down
exhibited by the mobility of an overdamped particle moving in a
periodic potential and in an inhomogeneous medium \cite{Mah}, the
noise induced order in one-dimensional map of the
Belousov-Zhabotinsky reaction \cite{Mat}, and the one-peak
structure in the MFPT of a bistable kinetic model driven by
correlated noises \cite{Xie} are akin to the NES phenomenon.
Previous theoretical papers analyzed NES phenomenon in systems
with metastable and unstable states of fixed potential or
periodically driving potential \cite{Man,Agu,Fia,Day,Wac}. However
the model of randomly switching metastable state is more realistic
when we describe the generation process of the carrier traps in
semiconductors. Let us consider one-dimensional overdamped
Brownian motion in randomly switching potential profile
\begin{eqnarray}
\frac{dx}{dt}=-\frac{dU\left(  x\right) }{dx} +a\eta\left(
t\right) +\xi\left( t\right)
.\label{Lang}%
\end{eqnarray}
where $x(t)$ is the displacement in time $t$, $\xi(t)$ is the
white Gaussian noise with zero mean and $\left\langle
\xi(t)\xi(t+\tau)\right\rangle =2D\delta(\tau)$, $\eta(t)$ is
Markovian dichotomous process which takes the values $\pm1$ with
the mean rate of switchings $\nu$. We assume that potential $U(x)$
has reflecting boundary at the point $x=0$ and absorbing boundary
at the point $x=b$ $(b>0)$. Exact results for MFPTs for Brownian
diffusion in switching potentials were first derived in
\cite{Hang}. Here we use equations from \cite{Hang} but different
conditions at the reflecting boundary \cite{Zur}. Thus following
ref.\cite{Hang} we obtain the coupled differential equations for
the MFPTs of our system (\ref{Lang})
\begin{eqnarray}
&&DT_{+}^{\prime\prime}+\left[ a-U^{\prime}\left(  x\right) \right]  T_{+}^{\prime}%
+\nu\left(  T_{-}-T_{+}\right) =-1, \nonumber\\
&&DT_{-}^{\prime\prime}-\left[ a+U^{\prime}\left(  x\right) \right]  T_{-}^{\prime}%
+\nu\left(  T_{+}-T_{-}\right) =-1.
 \label{Hang}
\end{eqnarray}
Here $T_{+}(x)$ and $T_{-}(x)$ are respectively the MFPTs for
positive $\eta(0)=+1$ and negative $\eta(0)=-1$ initial value of
dichotomous noise. In accordance with ref. \cite{Zur} the
conditions at reflecting boundary $x=0$ and absorbing boundary
$x=b$ are $T_{\pm}^{\prime}\left( 0\right) =0$ and $T_{\pm}\left(
b\right) =0.$ Let us introduce for convenience two new auxiliary
functions $T=(T_{+}+T_{-})/2$ and $\theta=(T_{+}-T_{-})/2$. So
that from Eqs.~(\ref{Hang}) we obtain
\begin{eqnarray}
&&DT^{\prime\prime}-U^{\prime}\left(  x\right)
T^{\prime}+a\theta^{\prime} =-1, \nonumber\\
&&D\theta^{\prime\prime}-U^{\prime}\left(  x\right)  \theta^{\prime}+aT^{\prime}%
-2\nu\theta =0.
\label{twoT}%
\end{eqnarray}
with the following boundary conditions: $T^{\prime}\left( 0\right)
=\theta^{\prime}\left(  0\right) =0$ and $T\left( b\right)
=\theta\left(  b\right)  =0$. By eliminating $T(x)$ from
Eqs.~(\ref{twoT}) we arrive at third-order linear differential
equation for $\theta (x)$
\begin{eqnarray}
\theta^{\prime\prime\prime}-\frac{2U^{\prime}\left(  x\right)
}{D}\theta^{\prime \prime}+\left[  \frac{U^{\prime 2}\left(
x\right) }{D^{2}}-\frac{U^{\prime\prime}\left( x\right)
}{D}-\gamma^{2}\right] \theta^{\prime}+\frac{2\nu U^{\prime}\left(
x\right) }{D^{2}}\theta=\frac{a}{D^{2}},
\label{theta}%
\end{eqnarray}
where $\gamma=\sqrt{a^{2}/D^{2}+2\nu /D}$. Let us assume that in
Eq.~(\ref{Lang}) $U(x)=k(L-x)\cdot 1(x-L)$, where $1(x)$ is step
function and $0<L<b$. As we can see from Fig.~1 we have a
metastable state for $\eta(t)=-1$ and an unstable state for
$\eta(t)=+1$.
\begin{figure}
[ptb]
\begin{center}
\includegraphics[height=2.0219in,width=2.0219in]%
{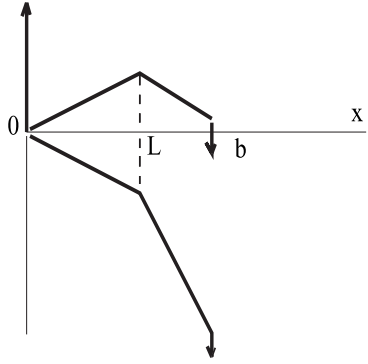}%
\caption{The piecewise linear potential $U(x)$ with metastable and
unstable states.}
\end{center}
\end{figure}
We will consider the MFPT $T_{+}(0)$, i. e. MFPT for initial
unstable state with starting position of Brownian particle at the
origin. In the absence of switchings and thermal diffusion the
dynamical MFPT $T_{+}(0)$ equals
\begin{eqnarray}
T_{+}(0)=\frac{L}{a}+\frac{b-L}{k+a}.
\label{dynam}%
\end{eqnarray}
We can solve Eqs.~(\ref{twoT}) and (\ref{theta}) separately for
regions $0<x<L$ and $L<x<b$. After some algebra we obtain finally
the MFPT $T_{+}(0)$ as
\begin{eqnarray}
T_{+}\left(  0\right)  =T\left(  0\right)  +\theta\left(  0\right)
=c_{1}\frac{\gamma^{2}D^{2}}{a^{2}}+c_{2}-\frac{a^{2}}{\gamma^{4}D^{3}}.
\label{T+}%
\end{eqnarray}
The unknown constants $c_{1}$ and $c_{2}$ can be determined from
the boundary conditions and the continuity conditions of the
functions $\theta(x)$, $\theta^{\prime}(x)$, $T(x)$,
$T^{\prime}(x)$ at the point $x=L$.

\section{Analysis of noise enhanced stability phenomenon conditions}

By expanding the Eq.~(\ref{T+}) in power series on $D$, for very
low intensity we obtain
\begin{eqnarray}
T_{+}\left(  0\right)  \simeq\frac{2L}{a}+\frac{\nu
L^{2}}{a^{2}}+\frac {b-L}{k}-\frac{\beta\left(  1-\beta\right)
}{2\nu}\left(  1-e^{-s}\right)  + \frac{D}{a^2}F(s,\beta,\omega)
\label{expand}
\end{eqnarray}
where
\begin{eqnarray}
&&F(s,\beta,\omega) =\frac{\beta ^{3}\left[ 2+s(1+\beta ^{2})\right]}{%
\left( 1+\beta \right) \left( 1-\beta ^{2}\right) }\cdot e^{-s}
+\frac{\beta \left(1-\beta ^{2}-2\beta ^{3}\right) }{2\left( 1-\beta ^{2}\right) }\left(%
1-e^{-s}\right) \nonumber\\
&&-\frac{5+\beta }{2\left( 1+\beta \right) }+2\omega \left(
\frac{1}{1-\beta ^{2}}-\frac{3}{\beta }\right) -\frac{2\omega
^{2}}{\beta ^{2}} \label{F}
\end{eqnarray}
and $\beta$, $s$ and $\omega$ are dimensionless parameters
\begin{eqnarray}
\beta=\frac{a}{k}, \quad \omega=\frac{\nu L}{k}, \quad s =
\frac{2\omega}{1-\beta^{2} }\left(  \frac{b}{L}-1\right).\nonumber
\end{eqnarray}
The sign of the function $F(s,\beta ,\omega )$ in
Eq.~(\ref{expand}) determines the condition to observe the NES
phenomenon in considered system. We can write this condition by
the following inequality
\begin{eqnarray}
F(s,\beta,\omega)>0.
\label{NES}%
\end{eqnarray}
Let us analyze the structure of NES phenomenon region on the plane
$(\beta,\omega)$. In the case of very slow switching
$\nu\rightarrow0$ $(\omega\rightarrow0,\;s\rightarrow0)$
Eq.~(\ref{NES}) takes the form
\begin{eqnarray}
\beta>0,802; \quad\omega<\frac{2\beta^{2}\left(  1-\beta\right)
-5\beta\left(  1-\beta^{2}\right)  ^{2}/2}{6\left(
1-\beta^{2}\right) ^{2}-2\beta\left(  1-\beta^{2}\right)
+\beta^{2}\left(  3\beta^{2}-1\right) \left(  b/L-1\right)}.
\label{regions}%
\end{eqnarray}
In the case of $\beta\simeq1$ we obtain from Eqs.~(\ref{F}),
(\ref{NES}) and (\ref{regions})
\begin{eqnarray}
\omega<\frac{1-\beta}{b/L-1}, \qquad\frac{1}{2}+\frac{5}{2}\left(
1-\beta\right)  <\omega<\frac{1}{2(1-\beta)}. \nonumber
\end{eqnarray}
In Fig.~2 are shown two NES regions (shaded areas) on the plane
$(\beta,\omega)$.
\begin{figure}
[ptb]
\begin{center}
\includegraphics[trim=-0.020199in -0.020199in 0.020199in 0.020199in,
height=2.0894in,width=2.0894in]%
{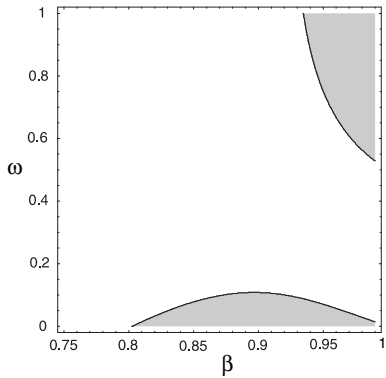}%
\caption{The shaded areas are the region of the plane
$(\omega,\beta)$ where the NES effect takes place.}
\end{center}
\end{figure}
As we can see from Fig.~2, the NES effect occurs only at the
values of $\beta$ near $1$, i.e. at very small steepness
$k-a=k(1-\beta)$ of the reverse potential barrier for the
metastable state (see Fig.~1). This NES region is different from
that obtained for the case of periodical dichotomous force
\cite{Agu}. On our opinion, this narrowing of NES phenomenon area
in comparison with the case of periodically driven metastable
states is due to the uncertainty of metastable state starting
time. The NES effect disappears when we choose the absorbing
boundary at the point $y=L$.

\section{Acceleration of diffusion in fluctuating periodic potentials}

The problem of diffusion in a periodic potential is related to a
rich variety of physical situations such as diffusion of atoms in
crystals, synchronization of oscillations, fluctuations of a
Josephson supercurrent and so on. An exact expression for
diffusion coefficient was independently obtained in the overdamped
limit by various methods for an arbitrary periodic potential and
for sinusoidal potential \cite{Fes}. Recently the mean velocity
and effective diffusion coefficient of Brownian particle moving in
a tilted periodic potential have been found in \cite{Lind}. The
case of diffusion in fast fluctuating periodic field was first
investigated in ref. \cite{Mal}, where the exact result for
diffusion coefficient in the sawtooth potential was derived. In
this section we generalize this result to the case of arbitrary
potential profiles (see also \cite{Alex}). We will find that the
problem of calculating an effective diffusion coefficient in
fluctuating periodic potential reduces to mean first-passage
problem similar to the case of fixed potential profile \cite{Wea}.
We consider an overdamped Brownian particle in fluctuating
periodic potential $U\left(x\right)$ whose dynamics is governed by
the Langevin equation
\begin{eqnarray}
\frac{dx}{dt}=-\frac{dU\left(  x\right)  }{dx}\cdot\zeta\left(
t\right) +\xi\left(  t\right)  ,
\label{Lang-2}%
\end{eqnarray}
where $x(t)$ is the displacement in time $t$, $\xi\left(  t\right)
$ and $\zeta\left(  t\right)  $ are statistically independent
Gaussian white noises with zero means and intensities $2D$ and
$2D_{\zeta}$ respectively. Further we assume that potential
$U\left(x\right)$ is even function with period $L$ and place the
origin in one of the potential minima. Following ref. \cite{Fes}
we determine the effective diffusion coefficient as the limit
\begin{eqnarray}
D_{eff}=\lim_{t\rightarrow\infty}\frac{\left\langle
x^{2}(t)\right\rangle }{2t}
\label{eff}%
\end{eqnarray}
because we have $\left\langle x\left(  t\right) \right\rangle =0$.
From Eq.~(\ref{Lang-2}) the Fokker-Planck equation for probability
density $W\left( x,t\right) $\footnote{We interpret the stochastic
differential equation (\ref{Lang-2}) in Stratonovich's sense.} is
\begin{eqnarray}
\frac{\partial W}{\partial t}=D_{\zeta}\frac{\partial}{\partial
x}U^{\prime }\left(  x\right)  \frac{\partial}{\partial
x}U^{\prime}\left(  x\right) W+D\frac{\partial^{2}W}{\partial
x^{2}}.
\label{Fokker}%
\end{eqnarray}
We can choose arbitrary initial condition for Eq.~(\ref{Fokker}),
since we look at the asymptotic behavior of mean square coordinate
of Brownian particles. We place therefore all Brownian particles
in the origin at $t=0$: $W\left( x,0\right)
=\delta\left(x\right)$. Because of periodicity of the potential,
the diffusion process can be coarsely conceived as consecutive
transitions of Brownian particle from points of potential minima
$x_{m}=mL$ to nearest neighboring points $x_{m\pm1}$. The
transition time represents the escape time over left or right
absorbing boundaries $x=x_{m\pm1}$ for particle starting from the
point $x=x_{m}$, i.e. the random first-passage time. Thus, we can
consider as in \cite{Lind}, the discrete process
\begin{eqnarray}
\tilde{x}(t)=\sum_{k=0}^{n(0,t)}q_{i},
\label{discrete}%
\end{eqnarray}
where $q_{i}$ are random increments of jumps with values $\pm L$
and $n(0,t)$ denotes the total number of jumps in the time
interval $\left( 0,t\right)  $. In the asymptotic limit
$t\rightarrow\infty$ the "fine structure" of a diffusion is
unimportant, and the random processes $x\left( t\right) $ and
$\tilde{x}(t)$ become statistically equivalent. In particular,
$\left\langle x^{2}\left(  t\right)  \right\rangle \simeq
\left\langle \tilde{x}^{2}(t)\right\rangle $. Further, the random
increments $q_{i}$ and the random times $\tau_{j}$ between jumps
are statistically independent of one another because of the
markovianity of $x\left( t\right)  $ and have the same probability
distributions $W\left(  q\right)  $ and $w\left(  \tau\right)  $
respectively. Because of the symmetry of potential
$U\left(x\right)$ the probability density $W\left(q\right)$ reads
\begin{eqnarray}
W\left(  q\right)  =\frac{1}{2}\left[  \delta\left(  q-L\right)
+\delta\left(  q+L\right)  \right] .
\label{increm}%
\end{eqnarray}
Let us calculate the second moment $\left\langle
\tilde{x}^{2}(t)\right\rangle $ from Eqs.~(\ref{discrete}),
(\ref{increm})
\begin{eqnarray}
\left\langle \tilde{x}^{2}(t)\right\rangle =
{\displaystyle\sum\limits_{n=0}^{\infty}} P_{n}\left(  t\right)
{\displaystyle\sum\limits_{k=0}^{n}}
{\displaystyle\sum\limits_{l=0}^{n}} \left\langle
q_{k}q_{l}\right\rangle =
{\displaystyle\sum\limits_{n=0}^{\infty}} P_{n}\left( t\right)
{\displaystyle\sum\limits_{k=0}^{n}} \left\langle
q_{k}^{2}\right\rangle =L^{2}\left\langle n(0,t)\right\rangle,
\label{sec-mom}
\end{eqnarray}
where $P_{n}\left(t\right)$ is the probability of $n$ jumps in the
time interval $\left(0,t\right)$. On the other hand in the limit
$t\rightarrow\infty$ we have that
\begin{eqnarray}
\left\langle n(0,t)\right\rangle \simeq\frac{t}{\tau},
\label{mean-num}
\end{eqnarray}
where $\tau=\left\langle \tau_{j}\right\rangle $ is the MFPT for
Brownian particle with initial position $x=0$ and absorbing
boundaries at $x=\pm L$. After substitution of
Eqs.~(\ref{sec-mom}), (\ref{mean-num}) in Eq.~(\ref{eff}) we
arrive finally at the exact result
\begin{eqnarray}
D_{eff}=\frac{L^{2}}{2\tau}, \label{newD}
\end{eqnarray}
which was first derived in ref. \cite{Wea} for fixed periodic
potential. In fluctuating periodic potentials therefore the
calculation of diffusion coefficient $D_{eff}$ reduces to mean
first-passage time problem. According to Eq.~(\ref{Fokker}) we
must solve the following equation with conjugated kinetic operator
\begin{eqnarray}
D\tau^{\prime\prime}\left(  x\right)  +D_{\zeta}U^{\prime}\left(
x\right) \frac{d}{dx}\left[ U^{\prime}\left(  x\right)
\tau^{\prime}\left(  x\right) \right] =-1 \label{MFPT}
\end{eqnarray}
and boundary conditions $\tau\left(  -L\right) =0$, $\tau\left(
L\right)  =0$. Then we must put $x=0$, i.e. find $\tau=\tau\left(
0\right) $. Solving Eq.~(\ref{MFPT}) with these boundary
conditions, we finally obtain the following exact formula for
effective diffusion coefficient of Brownian particle in fast
fluctuating periodic potential $U\left(x\right)$
\begin{eqnarray}
D_{eff}=D\left[
\frac{1}{L}\int_{0}^{L}\frac{dx}{\sqrt{1+D_{\zeta}\left[
U^{\prime}\left(  x\right)  \right]  ^{2}/D}}\right]  ^{-2}.
\label{Main}
\end{eqnarray}
As evident from Eq.~(\ref{Main}), $D_{eff}>D$ for an arbitrary
potential profile $U\left(  x\right)  $, therefore diffusion of
Brownian particles accelerates in comparison with the case
$U\left( x\right)  =0$. This result fully confirms the assumption
previously proposed in \cite{Mal}. We emphasize that the value of
diffusion constant does not depend on the height of potential
barriers, as for fixed potential \cite{Fes}, but it depends on its
gradient $U^{\prime}\left( x\right)$. We can explain the
phenomenon of diffusion acceleration directly from
Eq.~(\ref{newD}). The potential barriers change places through a
random modulation and Brownian particles move from point $x=0$ to
point $x=L$ more rapidly in comparison with free diffusion case,
i. e. in the average particles move downhill for the most part of
the distance. Let us consider particular shapes of potential
$U\left(x\right)$.
\\ (a) For the sawtooth profile $U\left(x\right) =
2E\left\vert x\right\vert /L$ at $\left\vert x\right\vert \leq
L/2$ we easily obtain the Malakhov's exact result \cite{Mal}
\begin{eqnarray}
D_{eff}=D+D_{\zeta}\frac{4E^{2}}{L^{2}}. \label{pila}
\end{eqnarray}
(b) For sinusoidal potential $U\left( x\right) =E\sin^{2}\left(
\pi x/L\right)$ from Eq.~(\ref{Main}) we have
\begin{eqnarray}
D_{eff}=\frac{\pi^{2}D\left(  1+\gamma^{2}\right)
}{4\mathbf{K}^{2}\left(
\gamma/\sqrt{1+\gamma^{2}}\right)  },\qquad\qquad\gamma=\frac{\pi E}{L}%
\sqrt{\frac{D_{\zeta}}{D}}, \label{sin}
\end{eqnarray}
where $\mathbf{K}\left(k\right)$ is the complete elliptic integral
of the first kind $\left( 0<k<1\right)$. We derive from
Eq.~(\ref{sin}) at small strength $D_{\zeta}$ of modulating white
noise $\left( \gamma\ll1\right)$ the following approximating
expression
\begin{eqnarray}
D_{eff}\simeq D+D_{\zeta}\frac{\pi^{2}E^{2}}{2L^{2}}. \nonumber
\end{eqnarray}
This formula coincides with approximate result \cite{Mal} obtained
on the coarse assumption of Gaussian probability density $W\left(
x,t\right) $ although the real probability density is multi-modal.
In opposite case $\gamma\gg1$ using the asymptotic formula for
elliptic integral
\begin{eqnarray}
\mathbf{K}\left(  k\right) \simeq\ln\frac{4}{\sqrt{1-k^{2}}}\qquad
\qquad\left( k\rightarrow1\right)  , \nonumber
\end{eqnarray}
we find from\ Eq.~(\ref{sin})
\begin{eqnarray}
D_{eff}\simeq\frac{D\left(  \pi\gamma\right)
^{2}}{4\ln^{2}\gamma}\sim \frac{D_{\zeta}}{\ln^{2}D_{\zeta}}.
\label{large}
\end{eqnarray}
According to Eq.~(\ref{large}) the effective diffusion coefficient
increases with intensity $D_{\zeta}$ of modulating noise but more
slowly than linear law (\ref{pila}).\\
(c) Finally for piecewise parabolic periodic potential profile
\begin{eqnarray}
U\left(  x\right)  =\left\{
\begin{array}
[c]{cc}%
8E\left(  x/L\right)  ^{2}, & \left\vert x\right\vert \leq L/4\\
E\left[  1-8\left(  x/L-1/2\right)  ^{2}\right]  , & L/4\leq
x\leq3L/4
\end{array}
\right. \nonumber
\end{eqnarray}
we get from the exact formula (\ref{Main}) the following result
\begin{eqnarray}
D_{eff}=\frac{Dm^{2}}{\ln^{2}\left(  m+\sqrt{1+m^{2}}\right)
},\qquad\qquad m=\frac{4E}{L}\sqrt{\frac{D_{\zeta}}{D}}.
\label{parabolic}
\end{eqnarray}
At comparatively small intensity $D_{\zeta}$ $\left( m\ll1\right)$
we obtain
\begin{eqnarray}
D_{eff}\simeq D+D_{\zeta}\frac{16E^{2}}{3L^{2}} \label{small}
\end{eqnarray}
that is quite similar to the formula obtained for sinusoidal
potential. Moreover, the dependence of effective diffusion
coefficient on large $D_{\zeta}$ is similar to the law
(\ref{large}) for sinusoidal potential profile.

\section{Acknowledgements}

This work has been supported by INTAS Grant 2001-0450, MIUR, INFM,
by Russian Foundation for Basic Research (project 02-02-17517),
and Federal Program "Scientific Schools of Russia" (project
1729.2003.2).

\end{document}